\documentclass[journal]{IEEEtran}
\usepackage{epsfig}
\usepackage{amsmath}
\usepackage{color}
\usepackage{booktabs}
\usepackage{graphicx}          

\newtheorem{lemma}{Lemma}
\newtheorem{theorem}{Theorem}

\newtheorem{remark}{Remark}
\newtheorem{definition}{Definition}

\def\be{\begin{equation}}
\def\ee{\end{equation}}
\def\bea{\begin{eqnarray}}
\def\eea{\end{eqnarray}}
\newcommand{\re}[1]{(\ref{#1})}
\def\ne{ \nonumber \\ }
\def\nn{ \nonumber }

\begin{document}
\title{
Fractional Barrier Lyapunov Functions\\
with Application to Learning Control
}
\author{
Mingxuan Sun
\thanks{
This work was supported by National Science Foundation of China (62073291).}
\thanks{
The author is with the College of Information Engineering, Zhejiang University of Technology,
Hangzhou, 310023, Zhejiang, China.
Email address: mxsun@zjut.edu.cn
 }} \maketitle

\begin{abstract}
Barrier Lyapunov functions are suitable for learning control designs, due to their feature of finite duration tracking.
This paper presents fractional barrier Lyapunov functions, provided
and compared with the conventional ones in
the error-constraint learning control designs.
Two error models are adopted and the desired compensation control approach is applied for a non-parametric design,
allowing two kinds of uncertainties involved in the error dynamics.
Theoretical results about existence of the solution and  convergence of the learning control schemes are presented.
It is shown that
fully-saturated learning algorithms play important role in assuring boundedness of the estimates,
by which
the error constraint objective can be achieved.
Moreover,
the robust technique is developed through modifying
the discontinuous action involved in the
learning control scheme
that yields the expected tracking performance in the presence of residual.
\end{abstract}


\section{Introduction} \label{sec.introduction}

Iterative learning control (ILC) features
time-variant-signal learning,
and exhibits the ability to improve the tracking performance of the closed-loop system undertaken,
when executing tasks in a repetitive manner.
Conventional ILC designs are carried out for {\it input learning} \cite{wang00},
while adaptive ILC schemes conduct {\it parameter learning}
for unknowns that involved in system dynamics \cite{sadegh90}.
An adaptive ILC control design,
 on the basis of
the certainty-equivalence principle,
underlines
time-varying parametrization
for the system model or its controller.
The conventional integral adaptation and the iterative updating can be combined for parameter adaptation \cite{qu02,tayebi09}. This
in certain sense clarifies
the connection of the learning methodology
with the conventional adaptive systems \cite{narendra89}.
The projection-based learning algorithms for assuring
bounded-estimates were reported in
\cite{sadegh90,marino09}.
The simple and direct versions are the partially-saturated
algorithms \cite{dixon02,xu04a} and the fully-saturated ones \cite{sun06}.
The mentioned input-constraint schemes lead to the boundedness
of the estimates themselves, not in the sense of $L_2$,
where the energy-like functional was shown to be suitable to the analysis of learning systems \cite{xu04a,xu06}.
Recently, in \cite{sun09}, the convergence performance was re-examined in the absence of residual.

Desired compensation control (DCC) designs were presented in \cite{sadegh90,yao09} where the unknowns
to be learnt are constants that can be dealt with by the conventional integral adaptation mechanisms.
As such, the parametrization for the control design is required.
As a unified way, this approach was also shown to be applicable for addressing the repetitive control problem of robotic systems \cite{sadegh90}.
The works reported in \cite{marino09,xu14}
are closely related,
where by assuming the desired dynamics, no direct parametrization is required.
It should be noted that in the above-mentioned, the learning process involves no residuals.
However, fewer works dealt with
the situation of residuals.
To tackling the norm-bounding uncertainties, the robust treatment with the use of the signum function is efficient so that the residuals can be avoided \cite{xu14,li16}.
The robustness improvement in case of nonzero approximation error
is made possible due
to the use of a deadzone modified Lyapunov functional in \cite{sun18}.

Besides the input constraint,
the other critical issue is the state/output constraint within a prescribed region, due to the finite-interval-operation feature of ILC.
The barrier Lyapunov function (BLF) based control designs
are helpful to the constraint purpose.
One of the pioneer works was reported in \cite{tee09}
who formally exploited the barrier-Lyapunov synthesis.
The published results are available but bits and pieces, and not received attentions they deserve,
addition to the suffers a few available barrier functions.
Logarithmic-barrier function is the typical one, appeared in many related works.
Other one is the atangent-barrier function \cite{xu13},
the first effort made for the constraint learning control design.
Note that both logarithmic- and atangent-barrier functions are transcendental functions.
For easy implementation, the barrier functions that give strong barrier actions but
need less online computation are wanted.
Moreover,
the design problem of the constraint learning control
in the presence of residuals
is still open.

In this paper, we suggest novel fractional BLFs to realize
the error-constraint learning control.
Our preliminary result of this effort was reported in \cite{sun14}, where we
proposed to use a particular fractional BLF.
Error models are convenient
for adaptive system design and analysis
\cite{narendra89}, where the error dynamics are representatives for a broad class of practical systems.
We shall show how the DCC approach is helpful to
handle uncertainties involved in error dynamics, whenever it is available.
Conventional DCC designs underline that the desired system dynamics are assumed.
We shall clarify why in our learning control designs,
the conventional parametrization can be avoided.
To this end,
there are two kinds of uncertainties to be handled.
One is time-varying but iteration independent variables, with the regressor one, and the other is the norm-bounded term, which will vanish as the error tends zero.
We shall establish the existence of solution and convergence of the learning control schemes.
This approach can be compared with the existing techniques.
The adaptive system requires parametrization, and
the robust technique needs norm-bounded functions.
Comparing with the existing related works, the contributions of this paper lie in:
i) fractional BLFs for learning control designs, in order to restrict the error variables in the closed-loop;
ii) non-parametric DCC-based design method for the error models, needing not applying the parametrization method; and
iii) learning control method in the presence of residuals, together with
fully-saturated learning algorithms assuring boundeness of the estimates.

\section{Logarithmic BLF Revisited}
\label{sec.Lblf}

We usually carry out the constraint control redesign with a BLF,
and a Lyapunov function, $V$,
is available
for the nominal system undertaken.
Here it is assumed that $0 \le V(0) < b_V$,
where $b_V > 0$ is a chosen barrier bound by designer.
The currently conventionally used logarithmic BLF (LBLF) is given by
\bea \label{eq.lbf1}
f_{\rm LI}(V) = \log \frac{b_V}{b_V - V}.
\eea

One may wish the BLF undertaken to satisfy
the infinite barrier property (IBP).

\begin{definition}
A BLF $f(V,b_V)$ is said to satisfy
the infinite barrier property, if
\[
\lim_{b_V \rightarrow \infty} f(V,b_V) = c V,
\]
with $c >0$ being a constant,
$V$ being a given Lyapunov function and $b_V$ indicating the barrier bound.
\end{definition}

Whenever we need not provide any  barrier, we can set the bound $b_V$ to be larger enough, according to this property.
Obviously, IBP fails for
the above-mentioned BLF, $f_{\rm LI}(V)$, because
$\lim_{b_V \rightarrow \infty} f_{\rm LI}(V) = 0$.
We suggest the following LBLF, for which IBP holds,
\bea \label{eq.lbf2}
f_{\rm LII}(V) = \log \frac{b_V e^{V}}{b_V - V}.
\eea
In \cite{xu13},
such IBP was examined for an arctangent barrier function.

We prefer the BLFs with the larger values
as well as larger derivatives,
which offer stronger barring action than those with the smaller values.
According to Tab. \ref{dBFs.L},
$f_{\rm LII}(V)$ and its derivative  are of larger values than
the values of $f_{\rm LI}(V)$ and its derivative, although both the second-order derivatives are equal.
Hence, for the constraint control design,
the function $f_{\rm LII}(V)$ is deserved to be suggested.
Here, we introduce the notation that $f_{\rm LI}(V) \preceq f_{\rm LII}(V)$, for denoting the characterized property.

\section{Fractional Barrier Lyapunov Functions}
\label{sec.fblf}

Novel barrier functions,
associated with useful properties, are presented here.
We shall illustrate how a fractional BLF (FBLF) is formed, when a Lyapunov function is available, and explain why they
are effective and efficient.

A typical fractional BLF, applied in the learning control design \cite{sun14}, is in the form of
\bea
f_{\rm FI}(V) = \frac{V}{b_V - V}.
\label{eq.fbf1}
\eea
It follows that,
by the inequality, ${\rm log} x \le x -1$, for $x \ge 0$,
\bea
f_{\rm LI}(V) \le f_{\rm FI}(V).
\label{eq.lbf.fbf}
\eea
In addition, by noting that $\frac{b}{b - x} \ge 1, $ for $0 \le x < b$,
\bea
\frac{d}{d V} f_{\rm LI}(V) \le \frac{d}{d V} f_{\rm FI}(V),
\label{eq.dlbf.dfbf}\\
\frac{d^2}{d^2 V} f_{\rm LI}(V) \le \frac{d^2}{d^2 V} f_{\rm FI}(V).
\label{eq.dlbf.ddfbf}
\eea

From \re{eq.lbf.fbf}-\re{eq.dlbf.ddfbf},
$f_{\rm FI}(V)$ and its derivatives are of larger values than
the values of $f_{\rm LI}(V)$ and its derivatives.
Hence, $f_{\rm LI}(V) \preceq f_{\rm FI}(V)$, and $f_{\rm FI}(V)$ is one FBLF
which deserves to be fully explored and understood.

In addition, IBP holds for the following fractional FBLFs:
\bea
f_{\rm FII}(V) &=& \frac{b_V V}{b_V - V},
\label{eq.fbf2}\\
f_{\rm FIII}(V) &=& \frac{b_V +1-V}{b_V - V} V,
\label{eq.fbf3}\\
f_{\rm FIV}(V) &=& \frac{2b_V -V}{b_V - V} V,
\label{eq.fbf4}
\eea
and
\bea
f_{\rm FV}(V) &=& \frac{b_V +1}{b_V - V} V.
\label{eq.fbf5}
\eea

\begin{table}
\caption{The 1st-order and 2nd-order derivatives of LBLFs}
\begin{center}
\begin{tabular}{cccc}
\toprule
                & LBLFs & 1st-order derivatives  & 2nd-order derivatives \\
\midrule
$f_{\rm LI}(V)$ & $\log \frac{b_V}{b_V - V}$ & $\frac{1}{b_V-V}$ & $\frac{1}{(b_V-V)^2}$  \\
$f_{\rm LII}(V)$ & $\log \frac{b_V e^{V}}{b_V - V}$ & $1+\frac{1}{b_V-V}$ & $\frac{1}{(b_V-V)^2}$ \\
\bottomrule
\end{tabular}
\end{center}
\label{dBFs.L}
\end{table}

\begin{table}
\caption{The 1st-order and 2nd-order derivatives of FBLFs}
\begin{center}
\begin{tabular}{cccc}
\toprule
                & FBLFs & 1st-order derivatives  & 2nd-order derivatives \\
\midrule
$f_{\rm FI}(V)$ & $\frac{V}{b_V - V}$ & $\frac{b_V}{(b_V-V)^2}$ & $\frac{b_V}{(b_V-V)^3}$ \\
$f_{\rm FII}(V)$ & $\frac{b_V V}{b_V - V}$ & $\frac{b^2_V}{(b_V-V)^2}$ & $\frac{b^2_V}{(b_V-V)^3}$ \\
$f_{\rm FIII}(V)$ & $\frac{b_V +1-V}{b_V - V} V$ & $1+\frac{b_V}{(b_V-V)^2}$ & $\frac{b_V}{(b_V-V)^3}$ \\
$f_{\rm FIV}(V)$ & $\frac{2b_V -V}{b_V - V} V$ & $1+\frac{b_V^2}{(b_V-V)^2}$ & $\frac{b_V^2}{(b_V-V)^3}$\\
$f_{\rm FV}(V)$ & $\frac{b_V +1}{b_V - V} V$ & $\frac{b_V(b_V +1)}{(b_V-V)^2}$ & $\frac{b_V(b_V +1)}{(b_V-V)^3}$ \\
\bottomrule
\end{tabular}
\end{center}
\label{dBFs}
\end{table}

The first- and second-order derivatives of the mentioned-above FBLFs are listed in Tab. \ref{dBFs}.
Comparing with $f_{\rm FIII}(V)$,
we obtain the following relationships:
\bea
f_{\rm FI}(V) & \preceq & f_{\rm FIII}(V), \label{eq.f1f3}\\
f_{\rm LII}(V) & \preceq & f_{\rm FIII}(V). \label{eq.l2f3}
\eea
In the situation that $b_V \le 1$ (requiring that $V \le 1$), we have
\bea
f_{\rm FII}(V) & \preceq & f_{\rm FIII}(V); \label{eq.f2f3}
\eea
in comparison with $f_{\rm FIV}(V)$,
\bea
f_{\rm FII}(V) & \preceq & f_{\rm FIV}(V). \label{eq.f2f4}
\eea
Moreover, in comparison with $f_{\rm FV}(V)$,
\bea
f_{\rm FI}(V) & \preceq & f_{\rm FV}(V), \label{eq.f1f5}\\
f_{\rm FII}(V) & \preceq & f_{\rm FV}(V), \label{eq.f2f5}\\
f_{\rm FIII}(V) & \preceq & f_{\rm FV}(V). \label{eq.f3f5}
\eea

\begin{remark}
{\it
The fundamental form of LBLFs is given by
\re{eq.lbf1},
and \re{eq.fbf1} represents the basic form of FBLFs.
We see that \re{eq.fbf1} is simpler in form
than \re{eq.lbf1},
besides the advantageous properties stated by \re{eq.lbf.fbf}-\re{eq.dlbf.ddfbf}.
FBLF \re{eq.fbf2} is obtained by modifying \re{eq.fbf1}.
It follows from \re{eq.fbf3} that
$f_{\rm FIII}(V) = V + f_{\rm FI}(V)$;
from
\re{eq.fbf4}, $f_{\rm FIV}(V) = V + f_{\rm FII}(V)$;
and from
\re{eq.fbf5}, $f_{\rm FV}(V) = f_{\rm FI}(V) + f_{\rm FII}(V)$.
These
FBLFs are useful for the control design with a fixed barrier bound.
IBP is a helpful property
such that FBLFs \re{eq.fbf2}-\re{eq.fbf5} are applicable
as the barrier bound is set to be large enough.
}\end{remark}

\section{Constraint Iterative Learning Control}

In our present work we apply the DCC approach
for the uncertainty compensation
for the specified error models, along with the constraint control designs.
By applying the DCC approach, the parametrization is not needed
and the desired dynamics is not involved in the design,
but the problem that emerges is how to handel the estimation for time-varying nonlinearities.

\subsection{Control objective}
\label{sec.problem}

We will begin with our discussion
about the error models.

{\it Error model I} The nonlinear error model is described by
\bea
\dot{e}_k
&=& f(e_k,t) + g(e_k,t) ( u_k + \delta w_k(t) + \theta(t) )
\label{keyeq.errorsys}
\eea
where
$t \in [0,T]$ and $k$ is the iteration index;
$x_k$ is the $n-$dimensional vector of the system state,
and $u_k$ is the $m$-dimensional vector of the control input,
$e_k = x_k - x_d$ represents the state error of the $k$th cycle,
with
$x_d$ being the desired trajectory given {\it a priori} on $[0,T]$;
$\delta w_k(t) (= w(x_k,t) - w(x_d,t))$,
$w(\cdot, \cdot)$ indicating the lumped uncertainty,
and $\theta(t) = w(x_d,t)$.
Both $f(\cdot,\cdot)$ and $g(\cdot,\cdot)$
represent the nonlinearities of the error system undertaken.

The expressions for $\delta w_k(t)$ and $\theta(t)$ in \re{keyeq.errorsys}
are due to the DCC approach.
For such error model, we face two kinds of uncertainties to be
coped with.
Here, $\theta(t)$ is time-varying but state-independent, with the regressor being 1.
It is
assumed
for dealing with $\delta w(x_k,x_d,t)$
that $\|\delta w_k(t)\| \le \rho(x_k,x_d,t)$,
and
$\rho(x_k,x_d,t)$ tends to zero, as $x_k$ tends to $x_d$.

\begin{remark}
{\it
The learning control designs, presented in this paper, have to tackle the problem arisen from the norm-bounded uncertainty,
besides time-varying parameter estimation.
Two typical cases are as follows:
i) As $w$ is continuously differentiable on $[0,T]$ for all $k$, $\rho(x_k,x_d,t) = l_w \|e_k\|$.
This implies a global Lipschitz condition.
The control design can apply the bound directly, when $l_w$ is known.
Improved one is to conduct  estimation for $l_w$.
And
 ii) $\rho(x_k,x_d,t) = \rho_e(x_k,x_d,t) \|e_k\| $,
a Lipschitz-like condition.
The bound
$\rho_e(x_k,x_d,t)$
is usually assumed to be known.
However,
throughout this paper, we do not assume such Lipschitz-like conditions.
}\end{remark}

Let the origin be an equilibrium
of the nominal system, due to that $f(0,t)=0, \forall t \in [0,T]$.
Assume that the origin of the nominal system is globally asymptotically stable.
Then there exists a continuous differential function, $V(e_k,t)$, such that
\bea
&& \alpha_1(\|e_k\|)  \le  V(e_k,t) \le
\alpha_2(\|e_k\|)\label{eq.alpha.1} \\
&&\frac{\partial V(e_k,t)}{\partial t} +
L_f V(e_k,t)
 \le  - \alpha(\|e_k\|)\label{eq.alpha.2}
\eea
where $\alpha_1(\cdot), \alpha_2(\cdot)$ and $\alpha(\cdot)$ are class
$K_\infty$ functions.

The control objective of this paper is to find $u_k$ such that the error $e_k$ converges to a neighborhood of the origin, as faithfully as possible on $[0,T]$,
as $k$ increases.
At the same time, $e_k$ is enforced within a pre-specified region,
for any $t \in [0,T]$ and for all $k$.
In order to achieve this objective,
the category of initial conditions are taken into account throughout this paper, that
the actual initial state is set to be the same as the desired one, i.e.,
$x_k(0) = x_d(0)$, for all $k$.

\begin{remark}
{\it
As $e_k(0) =0$, for all $k$,
$\alpha_1(e_k(0))=0$ and $\alpha_2(e_k(0))=0$,
implying that
$V(e_k(0),t) =0$, for any $t \in [0,T]$.
}\end{remark}

We will also address the control design problem for the following error model, based on the DCC approach,

{\it Error model II}
A simple form of the error model is as follows:
\bea
 \dot{e}_{k} &=& A e_k +
  b (u_k + \delta w_k(t) + \theta(t) )
\label{eq.varepsilon.ex}
 \eea
where both $A$ and $b$ are with appropriate dimensions.
The matrix $A$ is stable so that
there exists positive definite matrix $P$ such that
$A^T P + PA =-Q$, for the given positive definite matrix $Q$.

\begin{remark}
{\it
There are practical examples which are expressed with the presented error models.
Special models were undertaken,
and thus the results of this paper are applicable
for robotic systems \cite{dixon02}.
}\end{remark}

With the use of the proposed fractional BLFs,
the error-constraint learning control designs for both models I and II are carried out, respectively.
In addition,
the theoretical results about existence of the solutions,
stability and convergence of the closed-loop systems
are presented.

\subsection{Robust ILC with discontinuous action}

Using the given fractional BLF,
the control law applied for model I is proposed as
\bea
u_k &=& - \theta_k - \varsigma_{k}\label{eq.contlaw}\\
 \varsigma_k &=& \left \{
\begin{array}{lcl}
\frac{z_k}{\|z_k \|}\rho_k, &\mathrm{if} & z_k \neq 0 \\
  0, &\mathrm{if} & z_k = 0
\end{array} \right . \label{eq.varsigma}
\eea
with the following learning law used for the parameter adaptation, for each $t \in [0,T]$,
 \bea \theta_k(t) &=& {\rm sat}
(\theta_{k}^*(t)) \label{eq.Llaw.1}\\
 \theta_k^*(t) &=& {\rm sat}(\theta_{k-1}^*(t))
+ \gamma z_k(t)
\label{eq.Llaw.2}
 \eea
where $\gamma >0$,
$z_k = \frac{b_V^2}{(b_V - V_k)^2} L_g V_k $, and
the notation $V_k(t) = V(e_k,t)$, is used for
simplicity of presentation.
Here,
$\frac{b_V V_k}{b_V - V_k}$ is the BLF $f_{\rm FII}$ we suggest, and
$b_V$ is the bound on $V_k$ we wish to set up.

The following lemma is provided to aid the theoretical analysis.

\begin{lemma} For positive sequences $r_k$ and $s_k$,
if
\bea
r_k \le r_{k-1} - s_k
\label{kineq}
\eea
and $r_0$ is bounded,
then the sequences $r_k$ and $s_k$ are bounded for all $k$,
and
$ \lim \limits_{k \to \infty } s_k = 0$.
\label{keylem.0}
\end{lemma}

\begin{theorem}
The solution of the ILC system, consisting of the error system \re{keyeq.errorsys}, the control law \re{eq.contlaw}-\re{eq.varsigma}, together with the learning law \re{eq.Llaw.1}-\re{eq.Llaw.2},
exists on $[0,T]$ for all $k$.
Moreover,
the tracking error $e_k(t)$ converges to
zero uniformly on $[0,T]$, as $k \rightarrow \infty$.
\label{thm.v}
\end{theorem}

\begin{proof}
To begin with we appeal for
the existence theorem, from which
there exists $t_1, 0 < t_1 < T,$ such that the solution exists on the interval $[0,t_1)$.
Suppose that $[0,t_1)$ is
the maximal interval of existence of the solution, and it cannot be continued up.
With \re{eq.contlaw}-\re{eq.varsigma},
the derivative of the chosen BLF along \re{keyeq.errorsys} can be calculated as
\bea
\frac{d}{dt}\frac{b_V V_k}{b_V - V_k}
&\le&
-\frac{b_V^2}{(b_V - V_k(t))^2} \alpha_k \ne &&
  + z_k^T ( u_k + \delta w_k + \theta ) \ne
&=&-\frac{b_V^2}{(b_V - V_k)^2}
 \alpha_k
   + z_k^T \tilde{\theta}_k
 \label{eq.dV.exist}
\eea
where $\tilde{\theta}_k=\theta_k-\hat{\theta}_k$.
With \re{eq.Llaw.1}-\re{eq.Llaw.2}, we obtain
$ \gamma z_k^T \tilde{\theta}_k -
(\theta_k-
 \theta_{k-1})^T\tilde{\theta}_k
 =
( \theta - {\rm sat}(\theta_k^*) )^T (\theta_k^* - {\rm
sat}(\theta_k^*) ) $.
It follows from (20) in \cite{sun09} that
$
( \theta - {\rm sat}(\theta_k^*) )^T (\theta_k^* - {\rm
sat}(\theta_k^*) ) \le 0$. As such,
\bea
\gamma z_k^T \tilde{\theta}_k \le (\theta_k-
 \theta_{k-1})^T\tilde{\theta}_k
 \label{keyeq}
 \eea
by which
\re{eq.dV.exist} can be rewritten as
 \bea
 \frac{d}{dt}\frac{b_V V_k}{b_V - V_k}
&\le& -
\frac{b_V^2}{(b_V - V_k)^2}
 \alpha_k
   +\frac{1}{\gamma} (\theta_k-
 \theta_{k-1})^T\tilde{\theta}_k.
\nn
 \eea
Noting that $ V_k(0) = 0$ gives rise to
 \bea
 \frac{b_V V_k(t)}{b_V - V_k(t)}
&\le&
\frac{1}{\gamma} \int_0^t
   (\theta_k(s) -
 \theta_{k-1}(s) )^T \tilde{\theta}_k(s)
  ds
\label{eq.vb}
 \eea
for $t \le t_1$.

Since $\theta_k$ is uniformly bounded on $[0,t_1)$, the right-hand side of \re{eq.vb} is bounded, implying that the term
$\frac{b_V V_k}{b_V - V_k}$ is bounded.
In turn,
$V_k(t) \le b_V$ for $[0,t_1)$,
due to that $\frac{b_V}{b_V - V_k(t)} >1$.
It follows from \re{eq.alpha.1} that $e_k(t)$
is bounded on $[0,t_1)$.
This contradicts to that $[0,t_1)$ is the maximal interval of existence of solution. Hence, the solution can be continued up the boundary, and
the solution exists on $[0,T]$ for each $k$.

Eq. \re{eq.vb} holds on $[0,T]$, and
$V_k(t) \le b_V$ for all $t \in [0,T]$,
whenever $V_k(0) \le  b_V$.
By \re{eq.alpha.1} $e_k$ is bounded on $[0,T]$.
In turn,
$z_k$ is bounded on $[0,T]$, according to its definition,
and by \re{eq.contlaw} $u_k$ is bounded on $[0,T]$.

To proceed for establishing the convergence result,
let us choose the Lyapunov-Krasovskii functional
$ L_k(t) = \frac{b_V V_k(t)}{b_V - V_k(t)} + \frac{1}{2 \gamma} \int_{0}^t
 (\tilde{\theta}_k^T(s) \tilde{\theta}_k(s) ) ds$.
Using the equality
$ \tilde \theta_k^T \tilde \theta_k
-  \tilde{\theta}_{k-1}^T \tilde{\theta}_{k-1}
= - 2  \tilde \theta_k^T (\theta_k - \theta_{k-1})
 - (\theta_k - \theta_{k-1})^T (\theta_k - \theta_{k-1})
$,
the difference between $L_k(t)$ and $L_{k-1}(t)$ can be calculated as
 \bea && \Delta L_k(t)(= L_k(t) - L_{k-1}(t))
 \ne
 &\le& \frac{b_V V_k(0)}{b_V - V_k(0)}
 - \frac{b_V V_{k-1}(t)}{b_V - V_{k-1}(t)} \ne
&& - \int_{0}^t \frac{b_V^2 }{(b_V - V_k(s))^2}\alpha_k(s) ds
 + \int_{0}^t z_k^T(s) \tilde{\theta}_k(s) ds \ne &&
 - \frac{1}{\gamma} \int_{0}^{t}
 (\theta_k(s) - \theta_{k-1}(s))^T\tilde{\theta}_k(s) ds \ne &&
 - \frac{1}{2 \gamma} \int_{0}^{t}(\theta_k(s) -
 \theta_{k-1}(s) )^T(\theta_k(s) - \theta_{k-1}(s) ) ds.
 \nn
 \eea
It follows from \re{keyeq} that
\bea  \Delta L_k(t)
 &\le& \frac{b_V V_k(0)}{b_V - V_k(0)}
 - \frac{b_V V_{k-1}(t)}{b_V - V_{k-1}(t)} \ne &&
 - \int_{0}^t \frac{b_V^2}{(b_V - V_k(s) )^2}\alpha_k(s) ds \ne &&
 - \frac{1}{2 \gamma} \int_{0}^{t}(\theta_k(s) -
 \theta_{k-1}(s))^T \ne && (\theta_k(s) - \theta_{k-1}(s)) ds
 \nn
 \eea
which implies, by $V_k(0) = 0$,
\bea
\Delta L_k
\le  - \frac{b_V V_{k-1}}{b_V - V_{k-1}}
 \nn
 \eea
With the fact that $\frac{b_V}{b_V - V_k(s)} > 1$,
 \bea
\Delta L_k
\le - V_{k-1}.
 \label{keyeq.DL}
 \eea
$L_k(t)$
is monotonically decreasing for each $t \in [0,T]$.
$L_0$ is bounded, due to it
continuity on $[0,T]$, which renders the boundedness of
$L_k$.
As such, the limit of $L_k$ exits,
leading to
$\lim_{k \rightarrow \infty} V_k(t) =0$, for $t \in [0,T]$. In turn,
we conclude that from \re{eq.alpha.1},
$\lim_{k \rightarrow \infty} e_k(t) =0$, for $t \in [0,T]$.
These convergence results coincide to
Lemma \ref{keylem.0}.
It follows from \re{keyeq.errorsys} that $\dot e_k(t)$ is
bounded on $[0,T]$, implying that
$\lim_{k \rightarrow \infty} e_k(t) =0$ uniformly on $[0,T]$.
This completes the proof.
\end{proof}

\begin{remark}{\it
By Theorem \ref{thm.v}, $V_k$ is enforced to be within a pre-specified region, i.e., $V_k \le b_V$.
Although we make the constraint on $V_k$, not directly for $e_k$,
it follows from \re{eq.alpha.1} that
$\|e_k \| \le \alpha_1^{-1}(b_V)$
whenever $V_k \le b_V$.
}
\end{remark}

According to the result presented in Theorem \ref{thm.v},
the implementation for model II can be given.
The barrier Lyapunov function
$W_k = \frac{1}{2}
\frac{ b_e^2
  e_k^T P e_k } {b_e^2
  - e_k^T P e_k } $, a typical form of
of $f_{\rm FIII}$, is undertaken.
The ILC system undertaken consists of
the error dynamics \re{eq.varepsilon.ex},
and the learning algorithm
\re{eq.contlaw}-\re{eq.varsigma}, and \re{eq.Llaw.1}-\re{eq.Llaw.2},
by viewing
$z_k = \frac{b_e^2 e_k^T P b }{(b_e^2 - e_k^T P e_k)^2} $.
Under that $e_k^T(0) P e_k(0) < b_{e}^2$,
the derivative of the chosen function along \re{eq.varepsilon.ex} can be given as
\bea
\dot W_k
&=&
-\frac{1}{2}  \frac{b_e^2 e_k^T Q e_k}{(b_e^2 - e_k^T P e_k)^2}\ne
&&+  z_k (u_k + \delta w_k +  \theta )\ne
&\leq& -\frac{1}{2}  \frac{b_e^2 e_k^T Q e_k }{(b_e^2 - e_k^T P e_k)^2} + z_k^T \tilde{\theta}_k.
\nn
\eea
By (20) in \cite{sun09},
$\gamma z_k^T \tilde{\theta}_k - (\theta_k - \theta_{k-1})^T \tilde{\theta}_k\leq 0$. Then
\bea \dot W_k &\le&
-\frac{1}{2}
 \frac{b_e^2 e_k^T Q e_k }{(b_e^2 - e_k^T P e_k)^2} + \frac{1}{\gamma}(\theta_k - \theta_{k-1})^T \tilde{\theta}_k.
 \label{eq.dotL.key}
\eea
Since $\theta_k$ is uniformly bounded, the right-hand side of \re{eq.dotL.key} is bounded.
Hence, $\dot W_k(t)$ is bounded, leading to that
$W_k(t)$ is bounded, due to the boundedness of $W_k(0)$.
Since $\frac{b_e^2 } {b_e^2
  - e_k^T P e_k }
 > 1 $,
then $e_k^T P e_k < b_{e}^2$,
which renders that $e_k(t)$ is bounded.
Hence, the solution can be continued up to the boundary, and the solution of the ILC system exists on $[0,T]$ for each $k$.
In turn, $z_k$ is bounded according to its definition, and by \re{eq.contlaw} $u_k$ is bounded.

To proceed for establishing the convergence,
we appeal for the following relationship,
\bea
W_k(t)
&\leq& - \int_{0}^t \frac{b_e^2 e_k^T(s) Qe_k(s)}{(b_e^2 - e_k^T(s) P e_k(s))^2} ds \ne
&& + \int_{0}^t  z_k^T(s) \tilde{\theta}_k(s) ds
\nn
\eea
where the condition $e_k(0)=0$ is used.
The deference between $L_{k}(t)$ and $L_{k-1}(t)$
along \re{eq.varepsilon.ex} can be written as
\bea
 && \Delta L_{k}(t) \ne
&\le& - W_{k-1}(t)
-\frac{1}{2}
 \int_{0}^t
   \frac{b_e^2e_k^T(s) Q e_k(s) }{(b_e^2 - e_k^T(s) P e_k(s))^2} ds\ne
&&+ \int_{0}^t z_k^T(s) \tilde{\theta}_k(s) ds \ne &&
 - \frac{1}{\gamma} \int_{0}^t \tilde \theta_k^T(s) (\theta_{k}(s) - \theta_{k-1}(s)) ds \ne &&
-\frac{1}{2 \gamma} \int_{0}^{t} (\theta_k(s) - \theta_{k-1}(s) )^T (\theta_{k}(s) - \theta_{k-1}(s) ) ds.
\nn
\eea
With the learning law, we obtain
\bea \Delta L_{k}(t)
&\le&
- W_{k-1}(t)
 - \frac{1}{2}
 \int_{0}^t
  \frac{b_e^2 e_k^T(s) Q e_k(s)}{(b_e^2 - e_k^T(s) P e_k(s) )^2} ds \ne &&
-\frac{1}{2 \gamma} \int_{0}^{t} (\theta_k(s) - \theta_{k-1}(s) )^T \ne && (\theta_{k}(s) - \theta_{k-1}(s) ) ds
 \nn
\eea
implying that
\bea \label{keyeq.dL.e}
\Delta L_{k} &\le& - W_{k-1}.
 \eea
Note that
$\frac{b_e^2} {b_e^2 - e_k^T P e_k} > 1$,
as $b_e^2 - e_k^T P e_k >0$.
It follows from \re{keyeq.dL.e} that
\bea
\Delta L_{k} &\le& - \frac{1}{2}
e_{k-1}^T P e_{k-1}.
 \nn
\eea
$L_0$  is bounded on $[0,T]$, due to its continuity.
By Lemma \ref{keylem.0} and the boundedness of $\dot e_k$,
the uniform convergence of
$e_k$ on $[0,T]$ can be established.

\section{Robust ILC with continuous action}
\label{sec.cilc.r}

The control law \re{eq.contlaw} involves the dis-continuous term.
We modify it to tackle the issue, by viewing
\bea \label{eq.varsigma.r}
\varsigma_k &=& \frac{\mu_k}{\|\mu_k \| + \epsilon} \rho_k
\eea
where $\epsilon>0$, $\mu_k = z_k \rho_k $, and
$z_k^T = \frac{b_V(b_V+1) }{(b_V - V_k )^2} L_gV_k$.
The same form of learning law \re{eq.Llaw.1}-\re{eq.Llaw.2} is applied.
Here, for constraint we choose
BLF $f_{\rm FIV}
=\frac{(b_V +1) V_k}{b_V - V_k}$,
and $b_V$ is the bound on $V_k$.

The following technical lemma is helpful for finalizing the performance analysis to be presented.

\begin{lemma} Given the sequence $d_k$, suppose that for positive sequences $r_k$ and $s_k$,
\bea
r_k \le r_{k-1} - s_k + d_k.
\label{kineq}
\eea
with $r_0$ being bounded,
and both satisfy that $s_k$ tends to zero
whenever $r_k$ does.
Then, i)
 $s_k$ is bounded for all $k$, and
$ \limsup \limits_{k \to \infty } s_k \le \bar d$,
as $d_k$ satisfies that $|d_k| \le \bar d$, for all $k$; and
ii)
$ \lim \limits_{k \to \infty }
s_k = 0$,
as $ \lim \limits_{k \to \infty } d_k = 0$.
\label{keylem}
\end{lemma}
\begin{proof} We refer to \cite{sun22} for the proof.
\end{proof}

\begin{theorem} \label{thm.ailc}
The solution of the modified ILC system
exists on $[0,T]$ for all $k$.
Moreover,
the tracking error $e_k(t)$ converges to a neighborhood of the origin,
with the radius proportional to the given $\epsilon$, on $[0,T]$, as $k \rightarrow \infty$.
\end{theorem}
\begin{proof}
At first, we assume that $[0,t_1)$, $0 < t_1 < T$, is
the maximal interval of existence of the solution. We calculate the derivative of the chosen BLF,
which satisfies
\bea
\frac{d}{dt} \frac{(b_V +1) V_k}{b_V - V_k}
&\leq& - \frac{b_V(b_V +1) }{(b_V - V_k)^2} \alpha_k \ne &&+z_k^T(u_k + \delta w_k + \theta ). \nn
\eea
Applying the modified control law,
\bea
&&z_k^T (u_k + \delta w_k + \theta ) \ne
&\leq& z_k^T \left ( \tilde\theta_k - \frac{\mu_k}{\|\mu_k \| + \epsilon} \rho_k \right ) + \|z_k\| \rho_k \ne
&\leq& z_k^T \tilde\theta_k +\frac{\epsilon\|\mu_k \|}{\|\mu_k \| + \epsilon} \ne
&\leq&  z_k^T \tilde\theta_k + \epsilon \nn
\eea
which leads to
\bea
\frac{d}{dt} \frac{(b_V +1) V_k}{b_V - V_k}
&\leq& -\frac{b_V(b_V +1) }{(b_V - V_k)^2} \alpha_k \ne && + z_k^T \tilde\theta_k + \epsilon. \label{keyeq.df}
\eea
With the modified learning law, \re{keyeq.df} can be rewritten as
\bea
 \frac{d}{dt} \frac{(b_V +1) V_k}{b_V - V_k} &\le& - \frac{b_V(b_V +1)}{(b_V - V_k)^2} \alpha_k  \ne
 && +\frac{1}{\gamma}  (\theta_k-
 \theta_{k-1})^T\tilde{\theta}_k + \epsilon \ne
 &\le&
 \frac{1}{\gamma}  (\theta_k-
 \theta_{k-1})^T\tilde{\theta}_k + \epsilon \nn
\eea
implying that, by $V_k(0) = 0$,
\bea
  \frac{(b_V +1) V_k(t)}{b_V - V_k(t)}
 \le
 \frac{1}{\gamma} \int_0^t (\theta_k(s) -
 \theta_{k-1}(s) )^T\tilde{\theta}_k(s) ds+ \epsilon t \label{keyeq.df.1}
\eea
for $t \le t_1$.

Since $\theta_k$ is uniformly bounded on $[0,t_1)$, the right-hand side of \re{keyeq.df.1} is bounded.
Note that $\frac{b_V +1} {b_V - V_k(t)} >1$,
assuring  that $V_k(t)\leq b_V$. In turn, $e_k(t)$ is bounded on $[0, t_1)$, which contradicts to the assumption that $[0,t_1)$ is the maximal interval of existence of solution. Hence, the solution of the modified ILC system exists on $[0,T]$ for all $k$.

For the convergence analysis,
we choose the barrier Lyapunov-Krasovskii functional,
$L_k(t) = \frac{(b_V +1) V_k(t)}{b_V - V_k(t)} + \frac{1}{2 \gamma} \int_{0}^t \tilde{\theta}_k^T(s) \tilde{\theta}_k(s) ds$.
It follows from \re{keyeq.df} that
\bea
\frac{(b_V +1) V_k(t)}{b_V - V_k(t)}
&\leq& \frac{(b_V +1) V_k(0)}{b_V - V_k(0)}  \ne && - \int_{0}^t\frac{b_V(b_V +1)}{(b_V - V_k(s))^2} \alpha_k(s) ds  \ne
&& + \int_{0}^t z_k^T(s) \tilde \theta_{k}(s)  ds
+ \epsilon t. \nn
\eea
Then,
the difference of $L_k$ and $L_{k-1}$
can be derived, which satisfies
\bea
\Delta L_k(t) &\leq& \frac{b_V+1 }{b_V - V_k(0)}V_k(0)- \frac{b_V+1}{b_V - V_{k-1}(t)}V_{k-1}(t)\ne
 &&- \int_{0}^t \frac{b_V(b_V +1)}{(b_V - V_k(s))^2}\alpha_k(s) ds
 - \frac{1}{2 \gamma} \int_{0}^{t} (\theta_k(s) \ne && - \theta_{k-1}(s) )^T(\theta_k(s) - \theta_{k-1}(s) ) ds \ne &&
 + \epsilon T. \nn
\eea
Since $V_k(0)=0$, then
\bea
\Delta L_k \leq - \frac{b_V+1}{b_V - V_{k-1}}V_{k-1} +\epsilon T. \nn
\eea
According to Lemma \ref{keylem}, we can conclude that
\bea
\overline{\rm lim}_{k\rightarrow\infty} \frac{b_V  V_{k}}{b_V - V_{k}} \leq \epsilon T. \nn
\eea
Due to $\frac{b_V }{b_V - V_{k}} \ge 1$, we have
\bea
\overline{\rm lim}_{k \rightarrow \infty}  V_k \le \epsilon T. \nn
\eea
Hence, using \re{eq.alpha.1},
$ \overline{\rm lim}_{k \rightarrow \infty} \|e_k\| \le \alpha_1^{-1} (\epsilon T )$.
This completes the proof. \end{proof}

The implementation can be made for
the ILC system,
by viewing
$z_k^T =\frac{b^2_e(b^2_e+1) }{(b^2_e - e_k^T P e_k )^2}e_k^T P b$, where
we choose the barrier Lyapunov function
$ W_k =
\frac{1}{2}\frac{( b^2_e+1) e_k^T P e_k}{b^2_e - e_k^T P e_k}$, a typical form of $f_{\rm FIV}$.
Assume that the interval $[0,t_1), 0 < t_1 < T$, is
the maximal interval of existence of the solution.
The derivative of the chosen function can be calculated as
\bea
\dot W_k &=&
-\frac{1}{2} e_k^TQe_k \frac{b^2_e(b^2_e+1) }{(b^2_e - e_k^TPe_k )^2}\ne
&&
+z_k^T (u_k + \delta w_k +  \theta ).\nn
 \eea
Similar derivations to arrive at \re{keyeq.df.1},
we obtain
\bea
\dot W_k &\leq& - \frac{1}{2}e_k^TQe_k \frac{b^2_e(b^2_e+1)}{(b^2_e - e_k^TPe_k )^2} \ne
&&+\frac{1}{\gamma}  (\theta_k-\theta_{k-1})^T\tilde{\theta}_k + \epsilon
 \nn
\eea
which gives rise to
\bea
\dot W_k &\leq& \frac{1}{\gamma}  (\theta_k-\theta_{k-1})^T\tilde{\theta}_k + \epsilon.
 \label{eq.dotL.key.r}
\eea
Since $\theta_k$ is uniformly bounded on $[0,t_1)$, the right-hand side of \re{eq.dotL.key.r} is bounded.
Hence, $\dot W_k(t)$ is bounded, implying that
$W_k(t)$ is bounded, due to the boundedness of $W_k(0)$.
Since $\frac{b^2_e }{b^2_e -e_{k}^T Pe_{k}} > 1$,
$e_k^T P e_k < b_{e}^2$ for $[0,t_1)$.
In turn, $e_k$
is bounded on $[0,t_1)$.
This contradicts that $[0,t_1)$ is the maximal interval of existence of solution, and
the solution of the modified ILC system exists on $[0,T]$ for each $k$.

With the chosen BLF,
the deference between $L_{k+1}(t)$ and $L_{k}(t)$ is calculated, which satisfies, by $e_k(0)=0$,
\bea
 \Delta L_k(t)
&\leq& - W_{k-1}(t) \ne &&
- \frac{1}{2} \int_{0}^t e_k^T(s)Qe_k(s) \frac{b^2_e(b^2_e+1) }{(b^2_e - e_k^T(s)Pe_k(s) )^2} ds \ne
&& - \frac{1}{2 \gamma}\int_{0}^t (\theta_k(s) - \theta_{k-1}(s) )^T(\theta_k(s) -\theta_{k-1}(s) ) ds \ne &&
+ \epsilon T
\nn
\eea
implying that
\bea
\Delta L_k \leq - W_{k-1} + \epsilon T. \label{keyeq.delL.e}
\eea
It follows from \re{keyeq.delL.e} that, by Lemma \ref{keylem},
\bea
 \overline{\rm lim}_{k\rightarrow \infty} \frac{b^2_e e_{k}^T P e_{k} }{b^2_e -e_{k}^TPe_{k}} \leq 2 \epsilon T. \nn
\eea
Using the fact $ \frac{b^2_e }{b^2_e - e_{k}^T} > 1$,
we obtain
\bea
 \overline{\rm lim}_{k\rightarrow \infty}e_{k}^T P e_{k} \leq 2\epsilon T. \nn
\eea
By the convergence result of $e_{k}^TPe_{k}$, $e_k$ converges to a neighborhood of the origin
on $[0,T]$, as $k \rightarrow \infty$,
with the radius proportional to  $\sqrt{2\epsilon T/\lambda_{\rm min}}$,
and $\lambda_{\rm min}$ being the minimum eigenvalue of the matrix $P$.

\begin{remark}{\it
It is seen that Lemma \ref{keylem} plays a crucial role in finalizing the analysis for an ILC system, in the presence of residuals.
It should be noted that
Lemma \ref{keylem} is applicable for various iterative  processes with residuals.
Lemma \ref{keylem.0} can be considered as a corollary of  Lemma \ref{keylem}, by setting $d_k =0 $ in \re{kineq}.
}\end{remark}

\begin{remark}{\it
The fully-saturated learning algorithm
\re{eq.Llaw.1}-\re{eq.Llaw.2}
ensure the uniform bounedness of the estimates, by which
the establishment for existence of solution
and the convergence assessment can be carried out.
}\end{remark}

\begin{remark}{\it
The constraint control technique
is more suitable for an ILC system,
because such a system runs over a finite interval, and the output has to be limited within the specified region.
By Theorem \ref{thm.v}, $e_k(t)$ is enforced to be within a pre-specified region, as $V_k(t) \le b_V$.
This in turn takes effect to restrict the state variables, as we shall show by the simulation result.
}
\end{remark}

\begin{remark}{\it
The control law \re{eq.contlaw}-\re{eq.varsigma}
may cause the chattering phenomenon, when implementing it.
We suggest to apply a modified control law, with the modified
term \re{eq.varsigma.r},
which makes the control smooth, and the chattering phenomenon can be avoided.
However, only the bounded-error convergence can be assured, with the convergence bound proportional to the $\epsilon$, an adjustable design parameter.
The applied DCC approach is different from the conventional robust techniques, since the norm-bound
$\rho_k$ decreases with respect to the tracking error,
and will approach zero, as the error tends to zero.
}\end{remark}

\section{Conclusion}

This paper has presented novel fractional barrier Lyapunov functions, and
the error-constraint ILC designs have been conducted for two error models.
The DCC approach has been shown applicable for developing the learning control schemes,
whenever the parametrization is not available.
Theoretical results about the existence of the solution and the convergence of the learning control algorithms have been presented.
It has been shown that
fully-saturated learning algorithms are effective to assure the boundedness of the estimates,
such that the objective of the error constraint can be achieved.
In addition,
the robust control technique, through modifying the discontinuous action,
has been shown to
yield the expected tracking performance
in the presence of residuals.

{\small

}

\end{document}